\documentclass[fleqn,usenatbib]{mnras}
\usepackage[T1]{fontenc}
\usepackage{ae,aecompl}
\usepackage{graphicx}
\usepackage{amsmath}
\usepackage{amssymb}
\usepackage{txfonts}
\usepackage{cleveref}
\usepackage{bm}
\usepackage{tensor}
\usepackage{mathtools}
\usepackage{upgreek}
\usepackage{xcolor}

\graphicspath{./}

\newcommand{\ltsima}{$\; \buildrel < \over \sim \;$}
\newcommand{\lsim}{\lower.5ex\hbox{\ltsima}}
\newcommand{\gtsima}{$\; \buildrel > \over \sim \;$}
\newcommand{\gsim}{\lower.5ex\hbox{\gtsima}}

\newcommand{\revision}[1]{\textcolor{black}{#1}}

\title[Cosmological Covariance of FRBs]
{Cosmological Covariance of Fast Radio Burst Dispersions}
\author[Reischke \& Hagstotz]
{
Robert Reischke\thanks{E-mail:  \href{mailto:reischke@astro.ruhr-uni-bochum.de}{reischke@astro.ruhr-uni-bochum.de}}$^{1}$ and Steffen Hagstotz\thanks{E-mail: \href{mailto:steffen.hagstotz@lmu.de}{steffen.hagstotz@lmu.de}}$^{2,3}$
\\
$^1$ Ruhr University Bochum, Faculty of Physics and Astronomy, Astronomical Institute (AIRUB),\\ \hspace{0.15cm} German Centre for Cosmological Lensing, 44780 Bochum, Germany
\\
$^2$  Universitäts-Sternwarte, Fakultät für Physik, Ludwig-Maximilians Universität München, \\
Scheinerstraße 1, D-81679 München, Germany
 \\
$^3$  Excellence Cluster ORIGINS, Boltzmannstraße 2, D-85748 Garching, Germany
}

\begin{document}
\onecolumn

\pagerange{\pageref{firstpage}--\pageref{lastpage}}
\pubyear{2022}
\maketitle
\label{firstpage}


\begin{abstract}
    The dispersion of fast radio bursts (FRBs) is a measure of the large-scale electron distribution. It enables measurements of cosmological parameters, especially of the expansion rate and the cosmic baryon fraction. The number of events is expected to increase dramatically over the coming years, and of particular interest are bursts with identified host galaxy and therefore redshift information.
    In this paper, we explore the covariance matrix of the dispersion measure (DM) of FRBs induced by the large-scale structure, as bursts from a similar direction on the sky are correlated by long wavelength modes of the electron distribution. We derive analytical expressions for the covariance matrix and examine the impact on parameter estimation from the FRB dispersion measure - redshift relation. The covariance also contains additional information that is missed by analysing the events individually. For future samples containing over $\sim300$ FRBs with host identification over the full sky, the covariance needs to be taken into account for unbiased inference, and the effect increases dramatically for smaller patches of the sky. \revision{Also forecasts must consider these effects as they would yield too optimistic parameter constraints. Our procedure can also be applied to the DM of the afterglow of Gamma Ray Bursts.}
\end{abstract}


\section{Introduction}
\label{sec:intro}
Fast radio bursts (FRBs) are very short transients lasting usually only a few milliseconds, with a frequency range from $\sim 100$ MHz to several GHz. 
The original pulse gets dispersed due to free electrons in the ionised intergalactic medium. This leads to a delayed arrival time of the pulse frequencies $\Delta t(\nu) \propto \nu^{-2}$, where the proportionality constant is called dispersion measure (DM)  \citep[e.g.][]{thornton_population_2013, petroff_real-time_2015, connor_non-cosmological_2016, champion_five_2016,chatterjee_direct_2017} and is related \revision{to} the column density of electrons along the line-of-sight to the FRB.

While the mechanism for the radio emission is still under debate,
their isotropic occurrence and large observed DM suggest an extragalactic origin for the vast majority of events \revision{\citep[even though some might also be galactic, see][]{andersen_bright_2020}}, so that the DM can be used to test the distribution of diffuse electrons in the large-scale structure (LSS). Several authors therefore proposed to use the DM inferred from FRBs as a cosmological probe, using either the average dispersion measure up to a given redshift \citep{zhou_fast_2014,walters_future_2018,hagstotz_new_2022,macquart_census_2020,wu_8_2022,james_measurement_2022} or the statistics of DM fluctuations \citep{masui_dispersion_2015,shirasaki_large-scale_2017,rafiei-ravandi_characterizing_2020,reischke_probing_2021,bhattacharya_fast_2020,takahashi_statistical_2021,rafiei-ravandi_chimefrb_2021,reischke_consistent_2022}. While the former requires host identification to acquire an independent redshift estimate, the latter can be done without it, as the homogeneous component can serve as a (noisy) estimate for the redshift. Angular statistics of the DM are formally very similar to cosmic shear since one is dealing with projections of cosmic fields. In this paper, however, we will focus on the homogeneous component of the DM, the so-called DM$-z$ relation, which can be employed in similar ways as supernovae Ia (SN Ia) measurements \citep[see e.g.][for the most recent results]{riess_comprehensive_2022, Brout:2022vxf}. The dispersion is used as a distance estimate and consequently as a probe of the geometry of the Universe. The total amplitude of the dispersion is also sensitive to the overall baryon content, the ionisation fraction and the Hubble constant. These are perfectly degenerate at the background level, so additional information about some of these quantities have to be considered to constrain the remaining one. A common choice is to adapt a prior on the baryon density coming from big bang nucleosynthesis as described in \citet{hagstotz_new_2022} in order to measure the Hubble parameter at late times.


Studies that employ FRBs to measure either the cosmic baryon density \citet{macquart_census_2020} or the Hubble constant \citet{hagstotz_new_2022} treat the individual bursts and their DM as independent. However, since the signal from an FRB travels through the large-scale structure (LSS), events within angular proximity on the sky become correlated. In this paper, we intend to fill this gap in current analyses and are concerned with deriving the covariance and its consequences for using the mean FRB dispersion for the inference of astrophysical and cosmological parameters. We emphasise that even though the observed signal does only depend on the cosmological background, the covariance itself is sensitive to fluctuations and therefore to perturbations charaterised by the 2-point correlation function of the electron distribution.


The paper is structured as follows: In \Cref{sec:theory} we summarise the theory of FRBs, the DM and derive the expression for the covariance matrix. \Cref{sec_disc} presents and discusses the results for a current sample of FRBs \citep{petroff_frbcat_2016} and the prospects for future analysis with FRBs. Finally, we summarise our findings in \Cref{sec:conclusion}. Throughout the paper we fix the cosmological parameters to a $\Lambda$CDM model with the best-fit values from the Planck mission \citet{Aghanim:2018eyx} and vary only one parameter for illustration, usually chosen to be the Hubble constant $H_0$.

\section{Testing the cosmological background with Fast Radio Bursts}
In this section we will review the basic theoretical framework of FRBs and how it is related to properties of the LSS. We will then derive main result of this paper, the covariance matrix for FRBs with host identification induced by the correlated LSS along nearby lines of sight.
\label{sec:theory}
\subsection{Dispersion Measure}
Cosmological tests using FRBs with host identification, that is with an independent redshift estimate, aim to fit the DM-$z$ diagram. The DM itself is estimated from the pulse's dispersion 
\begin{equation}
\label{eq:time_delay}
    \Delta t \propto \mathrm{DM}_\mathrm{tot}(\hat{\boldsymbol
    {x}}, z) \, \nu^{-2} \, ,
\end{equation}
defining the estimated DM of an FRB at the sky position $\hat{\boldsymbol{x}}$ and redshift $z$. Dispersion itself is caused by scattering with the free electrons along the line of sight. These electrons are either associated with the host halo, with the Milky Way, or with the large-scale structure (LSS). Therefore, the average total contribution can be split into three parts:
\begin{equation}
    \mathrm{DM}_\mathrm{tot}(\hat{\boldsymbol
    {x}}, z) = \mathrm{DM}_\mathrm{host}(z) + \mathrm{DM}_\mathrm{MW}(\hat{\boldsymbol
    {x}}) + \mathrm{DM}_\mathrm{LSS}(z,\hat{\boldsymbol
    {x}}) \; .
\end{equation}
Here the contribution from the Milky Way does not depend on redshift, since it is a local effect. Likewise the contribution from the host does not depend on the direction. The LSS contribution, however, depends both on redshift and direction, which will become important later on. Note that each of these contributions takes the form of a PDF with scatter around the mean values.

For this work, we will focus on the contribution from the LSS. We write explicitly 
\begin{equation}
\label{eq:DM_LSS}
    \mathrm{DM}_\mathrm{LSS}(\hat{\boldsymbol
    {x}},z) = \int_0^z \! n_\mathrm{e}(\hat{\boldsymbol
    {x}},z') \, f_\mathrm{IGM}(z') \, \frac{1+z'}{H(z')} \, \mathrm d z' \; ,
\end{equation}
where $n_\mathrm{e}(\hat{\boldsymbol
    {x}},z)$ is the {comoving} cosmic free electron density, $H(z) = H_0 E(z)$ is the Hubble function with the expansion function $E(z)$ and the Hubble constant $H_0$.  The overall DM is usually multiplied with the fraction $f_\mathrm{IGM}(z)$ of electrons in the IGM that are not bound in structures. For redshifts $z<3$ almost all baryons are ionised, it is thus useful to express the electron density by the number of baryons in the Universe:
\begin{align}
\label{eq:n_e}
    n_\mathrm{e}(\hat{\boldsymbol
    {x}},z) =  \chi_\mathrm{e} \frac{\rho_\mathrm{b}(\hat{\boldsymbol
    {x}}, z)}{m_\mathrm{p}} = \chi_\mathrm{e} \frac{\bar \rho_\mathrm{b}}{m_\mathrm{p}} \big( 1 + \delta_\mathrm{e} (\hat{\boldsymbol
    {x}}, z) )\;,
\end{align}
with the baryon density $\rho_\mathrm{b}$, the proton mass $m_\mathrm{p}$ and the electron fraction
\begin{align}
\label{eq:chi_e}
    \chi_\mathrm{e} &= Y_\mathrm{H} + \frac{1}{2} Y_\mathrm{He} \\
    & \approx 1 - \frac{1}{2} {Y}_\mathrm{He} \, ,
\end{align}
calculated from the primordial hydrogen and helium abundances $Y_\mathrm{H}$ and $Y_\mathrm{He}$. Here, we assume $Y_\mathrm{H} \approx 1 - Y_\mathrm{He}$ and $Y_\mathrm{He} = 0.24$, found to high precision both by CMB measurements \citep{Aghanim:2018eyx} and by spectroscopic observations of metal-poor gas clouds \citep{Aver:2015iza}.

The baryon number density in \Cref{eq:n_e} is commonly expanded around its background value $\bar \rho_\mathrm{b} / m_\mathrm{p}$ with the electron density contrast $\delta_\mathrm{e}$, whose mean vanishes by definition. Hence the DM is in principle a probe of the LSS by measuring DM statistics. This, however, requires a larger sample of FRBs than currently available. 

The electron fraction in the IGM in \Cref{eq:DM_LSS} is calculated by subtracting the fraction bound in stars, compact objects and the dense interstellar medium (ISM)
\begin{equation}
\label{eq:f_IGM}
    f_\mathrm{IGM}(z) = 1 - f_\star(z) - f_\mathrm{ISM}(z) \, .
\end{equation}
We compute\footnote{The code for the calculations is publicly available at \href{https://github.com/FRBs/FRB}{https://github.com/FRBs/FRB}, provided by \cite{macquart_census_2020}.} $f_\star$ and $f_\mathrm{ISM}$ using the estimates of star formation rate and ISM mass fraction from \cite{Fukugita:2004ee, Madau:2014bja}. We keep $f_\mathrm{IGM} = 0.84$ constant for the purposes of this paper.
Putting everything together, we write the DM -- redshift relation in \Cref{eq:DM_LSS} as
\begin{equation}
\label{eq:DM_LSS_v2}
    \mathrm{DM}_\mathrm{LSS}(\hat{\boldsymbol{x}},z) = \frac{3 \Omega_\mathrm{b0} H_0}{8 \pi G m_\mathrm{p}} \chi_\mathrm{e} \, f_\mathrm{IGM} \int_0^z \, \frac{1+z'}{E(z')}  \big(1+\delta_\mathrm{e}(\hat{\boldsymbol{x}},z')\big) \mathrm{d} z' \; ,
\end{equation}
with the dimensionless baryon density parameter $\Omega_\mathrm{b0}$ and the dimensionless expansion function $E(z) = H(z)/H_0$. Averaging \Cref{eq:DM_LSS_v2} provides the well known mean DM-redshift relation \citep{2003ApJ...598L..79I,2004MNRAS.348..999I, 2014ApJ...783L..35D}
\begin{equation}
\label{eq:DM_LSS_avg}
    \mathrm{DM}_\mathrm{LSS}(z) \coloneqq \langle\mathrm{DM}_\mathrm{LSS}(\hat{\boldsymbol{x}},z)\rangle = \frac{3 \Omega_\mathrm{b0} H_0}{8 \pi G m_\mathrm{p}} \chi_\mathrm{e} \, f_\mathrm{IGM} \int_0^z \, \frac{1+z'}{E(z')} \mathrm{d} z' \; .
\end{equation}
The measurement of FRBs together with a host redshift yields pairs $\{\mathrm{DM}_i, z_i\}$ and can be used to constrain any parameter from \Cref{eq:DM_LSS_avg} in addition to the cosmic expansion history.

\begin{figure}
    \centering
    \includegraphics[width=0.32\textwidth]{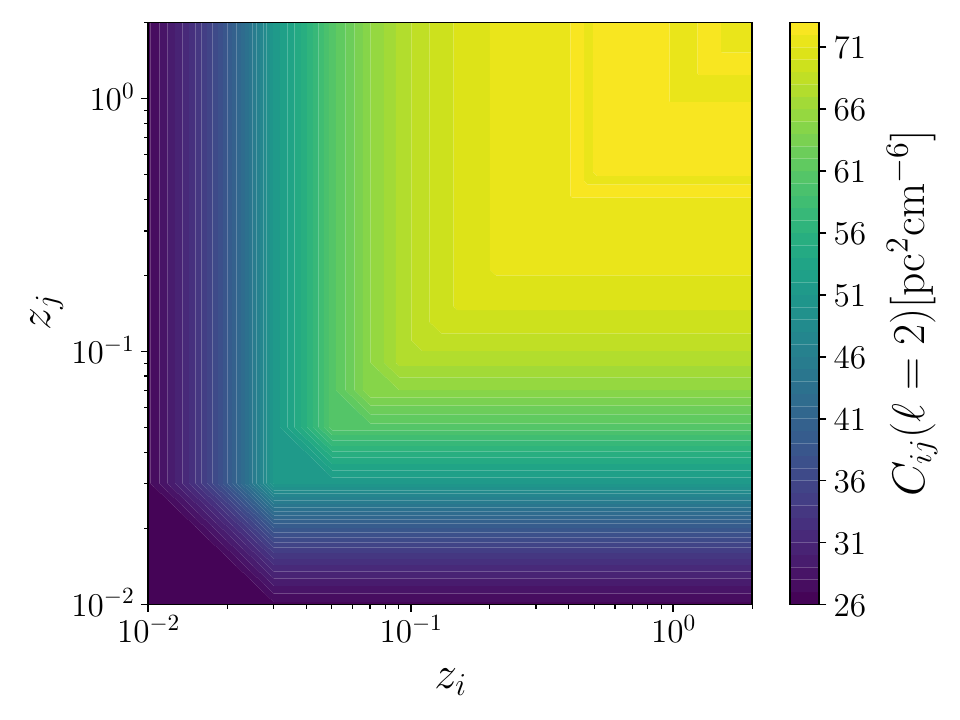}
    \includegraphics[width=0.32\textwidth]{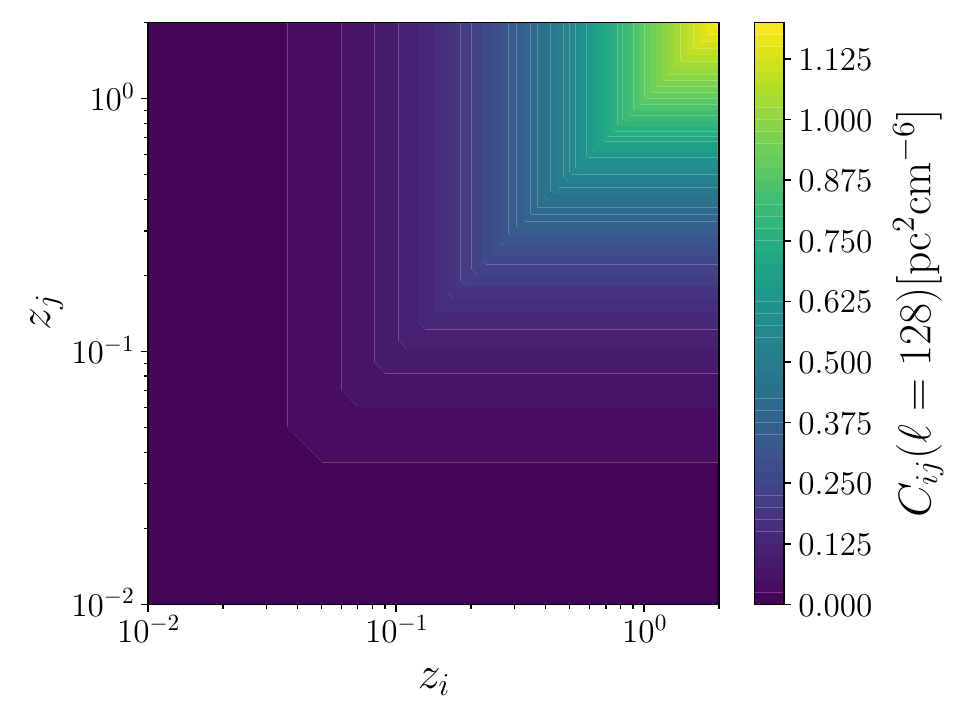}
    \includegraphics[width=0.32\textwidth]{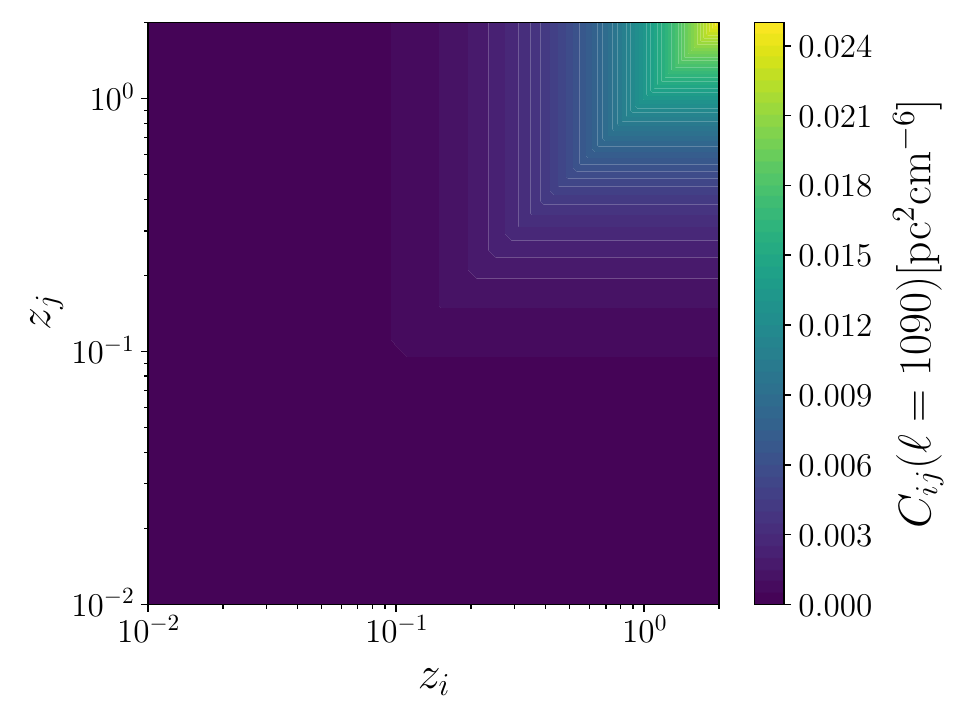}
    \caption{Angular power $C_{ij}(\ell)$ for different multipoles in the $(z_i,z_j)$-plane as defined in \Cref{eq:final_covariance}. Note that the colour scale changes and as well as the axis scaling of the rightmost plot.}
    \label{fig:C_ij_ell}
\end{figure}

\subsection{Covariance of the LSS component}
Observations of FRBs with host identification consist of a set of $N_\mathrm{FRB}$ measurements $\big\{\mathrm{DM}_i,\hat{\boldsymbol{x}}_i, z_i\big\}$, $i= 1,...,N_\mathrm{FRB}$, with the observed DM, the direction of the burst $\hat{\boldsymbol{x}}_i$ and its redshift. 
We are interested in the contribution to the covariance induced by the LSS between events labelled $i,j$:
\begin{equation}
   \mathrm{cov}_{ij} \coloneqq \left\langle \mathrm{DM}_\mathrm{LSS}(\hat{\boldsymbol{x}}_i,z_i)\mathrm{DM}_\mathrm{LSS}(\hat{\boldsymbol{x}}_j,z_j)    \right\rangle - \mathrm{DM}_\mathrm{LSS}(z_i)\mathrm{DM}_\mathrm{LSS}(z_j).
\end{equation}
Using \Cref{eq:DM_LSS_v2} and \Cref{eq:DM_LSS_avg} one finds
\begin{equation}
     \mathrm{cov}_{ij} = \int_0^{z_i}\mathrm{d}z'_i W_\mathrm{DM}(z'_i) \int_0^{z_j} \mathrm{d} z'_j \, W_\mathrm{DM}(z'_j)\left\langle\delta_\mathrm{e}(\hat{\boldsymbol{x}}_i,z'_i) \delta_\mathrm{e}(\hat{\boldsymbol{x}}_j,z'_j) \right\rangle\;,
\end{equation}
with the DM weight function:
\begin{equation}
    W_\mathrm{DM}(z) = \frac{3 \Omega_\mathrm{b0} H_0}{8 \pi G m_\mathrm{p}} \chi_\mathrm{e} \, f_\mathrm{IGM} \frac{1+z}{E(z)}\;.
\end{equation}
What is left to do is to work out the correlator in the integrand:
\begin{equation}
    \left\langle\delta_\mathrm{e}(\hat{\boldsymbol{x}}_i,z_i) \delta_\mathrm{e}(\hat{\boldsymbol{x}}_j,z_j) \right\rangle = \int\frac{\mathrm{d}^3k}{(2\pi)^3}\mathrm{e}^{\mathrm{i}\boldsymbol{k}\cdot(\boldsymbol{x}_i -\boldsymbol{x}_j)}P_e(k,z_i,z_j)\;,
\end{equation}
where we introduced the electron power spectrum and carried out the $k'$-integration. Expanding the exponential into plane waves yields:
\begin{align}
    \left\langle\delta_\mathrm{e}(\hat{\boldsymbol{x}}_i,z_i) \delta_\mathrm{e}(\hat{\boldsymbol{x}}_j,z_j) \right\rangle = & \; \frac{2}{\pi}\int k^2\mathrm{d}k\int\mathrm{d}\Omega_k P_e(k,z_i,z_j) \sum_{\ell,\ell'}\sum_{m,m'} \mathrm{i}^\ell (-\mathrm{i})^{\ell'} Y_{\ell m}({\hat{\boldsymbol{k}}}) Y^*_{\ell m}({\hat{\boldsymbol{x}_i}})j_\ell (k \chi_i) 
    Y^*_{\ell' m'}({\hat{\boldsymbol{k}}}) Y_{\ell' m'}({\hat{\boldsymbol{x}_j}})j_{\ell'} (k \chi_j)\\ 
    = & \; \frac{2}{\pi}\int k^2\mathrm{d}k P_e(k,z_i,z_j) \sum_{\ell}\sum_{m}Y^*_{\ell m}({\hat{\boldsymbol{x}_i}})j_\ell (k \chi_i) 
    Y_{\ell m}({\hat{\boldsymbol{x}_j}})j_{\ell} (k \chi_j)\\ 
    = & \; \frac{1}{2\pi^2}\sum_{\ell}(2\ell +1)\int k^2\mathrm{d}k P_e(k,z_i,z_j) j_\ell (k \chi_i) j_\ell (k \chi_j) 
    P_\ell(\cos \theta)\;.
    \end{align}
In the last step, we made use of the isotropy of cosmological fields and used 
\begin{equation}
    \sum_m Y_{\ell m}({\hat{\boldsymbol{x}_i}})Y^*_{\ell m}({\hat{\boldsymbol{x}_j}}) = \frac{2\ell + 1}{4\pi}P_\ell(\cos \theta)\;,
\end{equation}
with the Legendre polynomials $P_\ell(x)$ and we denote the angular separation between pairs of FRBs as $\hat{\boldsymbol{x}}_i\cdot \hat{\boldsymbol{x}}_j = \cos \theta$. Furthermore, $\boldsymbol{x} = (\hat{\boldsymbol{x}} \chi, \chi)$, where $\chi = \|{\boldsymbol{x}}\|$, with the comoving distance $\chi(z)$.
Thus, altogether, by using   $P_e(k,z_i,z_j) = \sqrt{P_e(k,z_i)P_e(k,z_j)} $ \revision{\citep[this approximation is accurate for all practical purposes and is discussed extensively in][for the case of cosmic shear, thus being very similar to the case studied here due to its broad weighting function along the line-of-sight]{Kitching:2016xcl, de_la_bella_unequal-time_2020}}, we arrive at
\begin{equation}
\begin{split}
\label{eq:final_covariance}
         \mathrm{cov}_{ij}(\cos \theta, z_i, z_j) = & \ \frac{1}{2\pi^2}\sum_{\ell}(2\ell +1) P_\ell(\hat{\boldsymbol{x}}_i\cdot \hat{\boldsymbol{x}}_j)\int k^2\mathrm{d}k\int_0^{z_i}\mathrm{d}z'_i W_\mathrm{DM}(z'_i) \sqrt{P_e(k,z'_i)} j_\ell (k \chi_i) \int_0^{z_j} \mathrm{d} z'_j \, W_\mathrm{DM}(z'_j)  \sqrt{P_e(k,z'_j)}j_\ell (k \chi_j) \\ 
        = & \ \sum_\ell \frac{2\ell+1}{4\pi}P_\ell(\cos \theta)C_{ij}(\ell)\;, 
        \end{split}
\end{equation}
which defines the angular power spectrum $C_{ij}(\ell)$ between the two fields $i$ and $j$. \revision{It should be noted that for large redshifts and large $\ell$, one can use the Limber approximation for the integrals over the Bessel function \citep{limber_analysis_1953,loverde_extended_2008}, which is accurate within $10\%$ for $\ell<30$ unless $z_i$ or $z_j$ are very small.}

\revision{To calculate the electron power spectrum, we use \texttt{HMX} \citep{mead_accurate_2015,mead_hydrodynamical_2020,troster_joint_2022} which uses a halo model based approaches to emulate power spectra of different species in the \texttt{BAHAMAS} simulations \citep{mccarthy_bahamas_2018}. Including e.g. free gas and cold dark matter. By using hydrostatic equilibrium and the appropriate mass conversion factors, these spectra can be converted into spectra of the electron pressure and density, allowing an accuracy of 15$\%$ on small scales. On large scales the electron bias approaches unity, i.e. the electron power spectrum is essentially given by the matter power spectrum.} In order to carry out the sum over $\ell$, we collect multipoles up to $\ell = 5\times 10^4$ on the diagonal and for the other entries take up to $\ell = 100/|\hat{\boldsymbol{x}}_i- \hat{\boldsymbol{x}}_j|$ into account.

\subsection{Remarks on parameter dependence of the covariance}

Since the covariance in \Cref{eq:final_covariance} depends on cosmological parameters, it contains additional information. There has been a long debate in the cosmological community whether it  is necessary to account for this dependence or not. Current LSS \citep[e.g.][]{asgari_kids-1000_2021,abbott_dark_2022} or CMB measurements \citep{aghanim_planck_2020} adjust the covariance interatively, that is they chose a fiducial cosmology, perform the inference for preliminary model parameters, update the covariance matrix to the preliminary best-fit model and start the inference again. This process is repeated until convergence is reached. \citet{carron_assumption_2013} discussed the assumption of a parameter (in)dependent covariance matrices when two-point statistics are used as the model and data, showing that the (Gaussian) covariance matrix never carries any independent information (as it is again just a product of two-point functions) and is rather a sign of non-Gaussian information. In \citet{reischke_variations_2017} the overall parameter dependence of the cosmic shear two-point covariance was investigated with analytic methods and ray-tracing simulations. This work was followed up by \citet{kodwani_effect_2019}, where the effect of a parameter dependent covariance matrix on the inference process with future LSS surveys was investigated and found to be negligible. However, one should keep in mind that these papers worked with averaged data and not simulated realisations of the data.
The situation studied in this paper is different since the average DM - redshift relation only contains information about the cosmological background, while the correlations in the data are induced by the perturbations characterised by the electron power spectrum. Therefore the covariance matrix contains additional information without any double-counting.

\begin{figure}
    \begin{center}     
    \includegraphics[width = .5\textwidth]{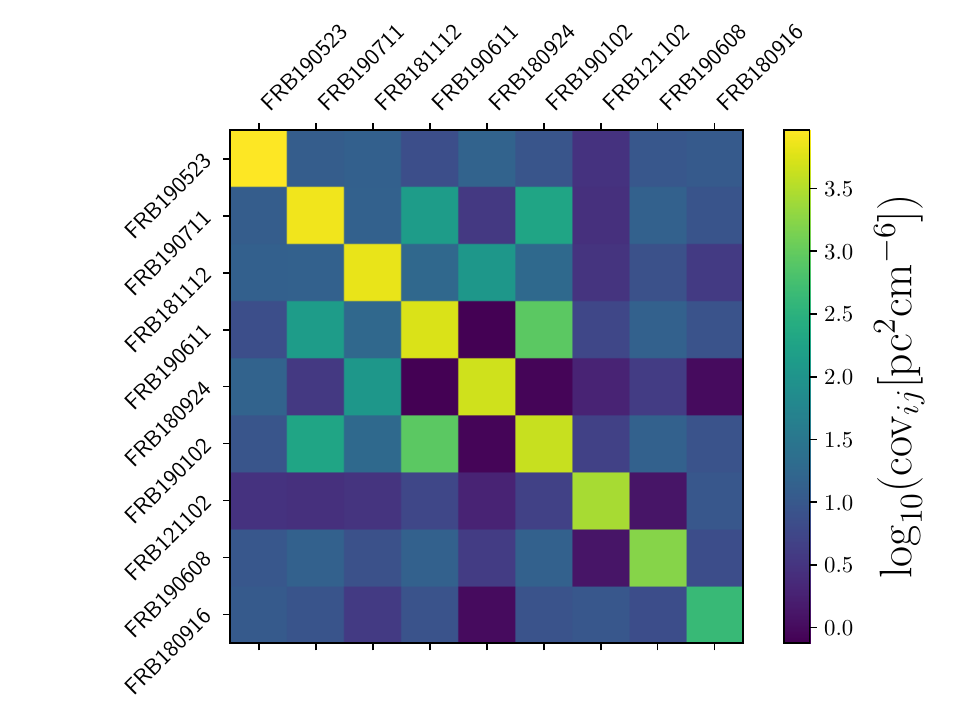}
    \includegraphics[width = .4\textwidth]{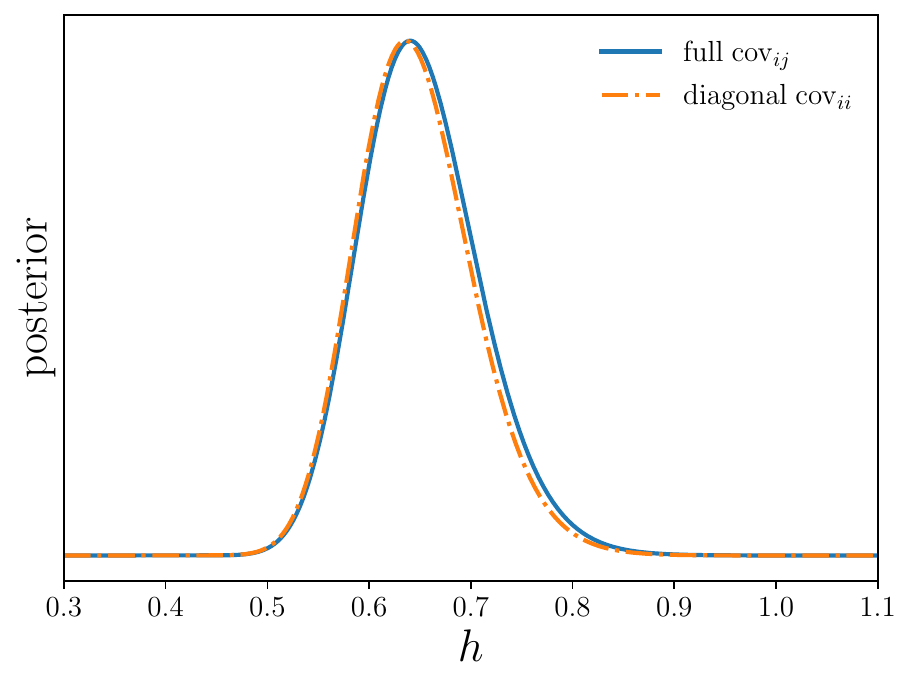}
    \end{center}
    \caption{\textbf{\textit{Left}}: Covariance matrix, \Cref{eq:final_covariance}, for the FRB catalogue \citep{petroff_frbcat_2016} with host identification. \textbf{\textit{Right}}: Posterior distribution of the Hubble constant (or other any amplitude of the DM), similar to the analysis carried out in \citet{hagstotz_new_2022}. The solid blue lines use the accurate covariance matrix, while the dashed orange lines only use the diagonal elements, i.e. the events are uncorrelated. Parameter dependence of the covariance does not change the results for this sample.}
    \label{fig:frb_cat}
\end{figure}

\section{Results and Discussion}
\label{sec_disc}
In this section we present the results for the covariance matrix. 
We start by discussing some intermediate results for the angular power spectra in the $(z_i,z_j)$-plane. \Cref{fig:C_ij_ell} shows the corresponding covariance for three different multipoles, $\ell = 2,\; 128, \; 1090$, from left to right. The colour bar encodes the covariance in redshift at these fixed angular scales $\ell \sim \theta^{-1}$. All covariances have a clear rectangular structure which stems from the integration bounds in \Cref{eq:final_covariance} reflecting the fact that the DM of two FRBs \revision{at $z_i$ and $z_j$} is only correlated for redshifts $z \leq\mathrm{min}(z_i,z_j)$. Furthermore, the structure of the covariance also shows that on larger angular scales the correlation is stronger at lower redshifts. This can be understood by the fact that the Bessel function $j_\ell(k \chi)$ peaks around $k \chi=\ell + 0.5$, thus small $\ell$ require small $\chi$ and hence $z$ to reach the peak of the power spectrum at $k\approx 0.01\;h^{-1}\mathrm{Mpc}$. Lastly, we also note that the variance obtained from \Cref{eq:final_covariance}, i.e. 
\begin{equation}
    \sigma^2_i = \sum_\ell (2\ell+1) C_{ii}(\ell)/(4\pi),
\end{equation}
agrees well with the results from the empirical formula presented in \citep{mcquinn_locating_2014,zhang_intergalactic_2020}:
\begin{equation}
    p(\Delta) \propto \Delta^{-\beta}\exp\left(\frac{(\Delta^{-\alpha} - C_0)^2}{2\alpha^2\sigma^2}\right)\;,
\end{equation}
with $\alpha = \beta = 3$, $\Delta = \mathrm{DM}_\mathrm{LSS}/\langle \mathrm{DM}_\mathrm{LSS} \rangle$ and the fitting values from $N$-body simulations presented in table 1 of \citet{zhang_intergalactic_2020}. At redshift $z=0.1$ we find 10 per-cent agreement with our analytical approach.

\subsection{Current Data}
We now turn to current data using all FRBs from the FRB catalogue \citep{petroff_frbcat_2016} with host identification. For illustrative purposes, we use them to fit the Hubble constant by putting a tight prior on the baryon density parameter $\Omega_\mathrm{b0}$. There are more events available at the time of writing \citep{james_measurement_2022}, but including a slightly larger sample does not affect the role of the covariance. In \citet{hagstotz_new_2022} the value of the physical density parameter, $\omega_\mathrm{b}= \Omega_\mathrm{b0} h^2$, as measured by big bang nucleosynthesis \citep{cooke_2018_bbn} was used, changing the overall scaling with $h$ slightly. With the approach followed here, one could think of the constraints just by looking at any linear amplitude parameter of the DM, \Cref{eq:DM_LSS_avg}. In \Cref{fig:frb_cat} we show the covariance matrix on the left for the 9 host-identified FRBs from the FRBCAT. Clearly the variance is largest for the highest redshifts, the cross-covariance, however, is largest between FRB190102 and FRB190611 which are in close proximity on the sky. Withal, the correlation coefficient is  below 0.2. The right panel shows the fit to the Hubble constant $H_0 = 100\;h\; \mathrm{km}\mathrm{s}^{-1}\mathrm{Mpc}^{-1}$ for these 9 FRBs. We assume a Gaussian likelihood
\begin{equation}
\label{eq:chi2}
    \chi^2 (\boldsymbol{\theta}) = \log\det \boldsymbol{C}(\boldsymbol{\theta}) + \left(\boldsymbol{d}-\boldsymbol{\mu}(\boldsymbol{\theta})\right)^T\boldsymbol{C}^{-1}(\boldsymbol{\theta}) \left(\boldsymbol{d}-\boldsymbol{\mu}(\boldsymbol{\theta})\right) \;,
\end{equation}
where we made the dependence on the parameters $\boldsymbol{\theta}$ explicit. The covariance consists out of three contributions
\begin{equation}
\label{eq:DM_covariance_all_contributions}
    \boldsymbol{C} = \boldsymbol{C}_\mathrm{LSS} + \boldsymbol{C}_\mathrm{MW} + \boldsymbol{C}_\mathrm{host}\;,
\end{equation}
and the components of $ \boldsymbol{C}_\mathrm{LSS} $ are given by \Cref{eq:final_covariance}, while we assume for the Milky Way $\boldsymbol{C}_\mathrm{MW} = \sigma^2_\mathrm{MW}\boldsymbol{\mathrm{I}}$, with $\sigma_\mathrm{MW} = 30\;\mathrm{pc}\;\mathrm{cm}^{-3}$ and the host $\boldsymbol{C}_\mathrm{host} = \sigma^2_\mathrm{host}\boldsymbol{\mathrm{I}}$, with $\sigma_\mathrm{host} = 50/(1+z)\;\mathrm{pc}\;\mathrm{cm}^{-3}$.

The results are shown on the right side of \Cref{fig:frb_cat}, where the solid blue line denotes the posterior using the full covariance matrix, while the assumption of independent events (taking only the diagonal of the covariance into account) leads to the dashed orange result. For the small sample size available right now, both approaches agree very well. In this case, the parameter dependence of the covariance is also still negligible. 
\revision{It will become larger as the sample size grows. This can be understood as follows: the effect of parameter dependence of the LSS component of the covariance matrix is of statistical nature. Thus more data points increase the signal-to-noise of the measurement of the errors, thus increasing its importance relative to the other components which only populate the diagonal elements (i.e. host and Milky Way contribution).}

\subsection{Future Data}
In order to illustrate when the proper treatment of correlated errors in the FRB dispersion becomes important, 
we now generate synthetic samples containing a total number of $N_\mathrm{FRB}$ FRBs distributed over redshift. For the redshift distribution, we assume a standard magnitude limited sample \citep[e.g.][]{reischke_probing_2021}:
\begin{equation}
    n(z) \propto z^2\exp(-z^\alpha)\;,
\end{equation}
with $\alpha = 5$. Next, we draw random positions for each FRB uniformly over patches in the sky with sky fractions $f_\mathrm{sky} = 1,\; 10^{-2}\; \mathrm{and}\; 10^{-3}$, so that the effective number density is $n = f_\mathrm{sky}^{-1}N_\mathrm{FRB}/(4\pi)$. For this sample, we calculate the covariance matrix \Cref{eq:final_covariance} of the LSS component which in turn yields the final covariance via \Cref{eq:DM_covariance_all_contributions}. We used this full covariance matrix to sample the $N_\mathrm{FRB}$ DM values for the generated events, completing the triples $\big\{\mathrm{DM}_i,\hat{\boldsymbol{x}}_i, z_i\big\}$ in our synthetic catalog.

\begin{figure}
    \centering
    \includegraphics[width = .5\textwidth]{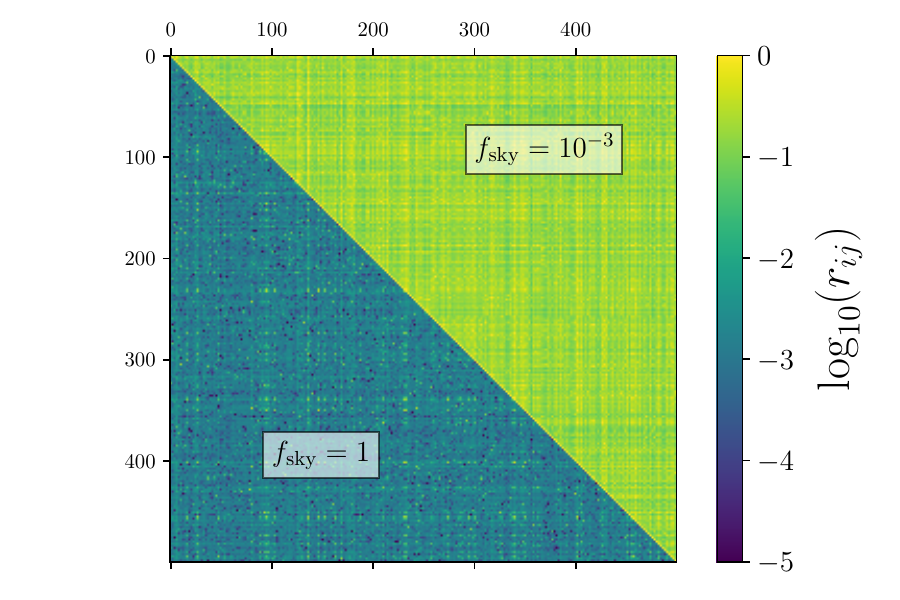}
    \caption{Correlation coefficient, $r_{ij} = \mathrm{cov}_{ij}/({\mathrm{cov}_{ii}\mathrm{cov}_{jj}})^{1/2}$, for 500 FRBs with host identification for a full sky (lower half) compared to the same sample only on a small subset $f_\mathrm{sky}= 10^{-3}$ of the sky (upper half), where the correlation of the data points becomes much stronger. The number of events corresponds to $n \approx 5 \times 10^{-3} \, \mathrm{deg}^{-2}$ for the full sky sample, and $n \approx 5 \, \mathrm{deg}^{-2}$ for the case of a small sky fraction.}
    \label{fig:covariance_forecast}
\end{figure}

\begin{figure}
    \centering
    \includegraphics[width = .9\textwidth]{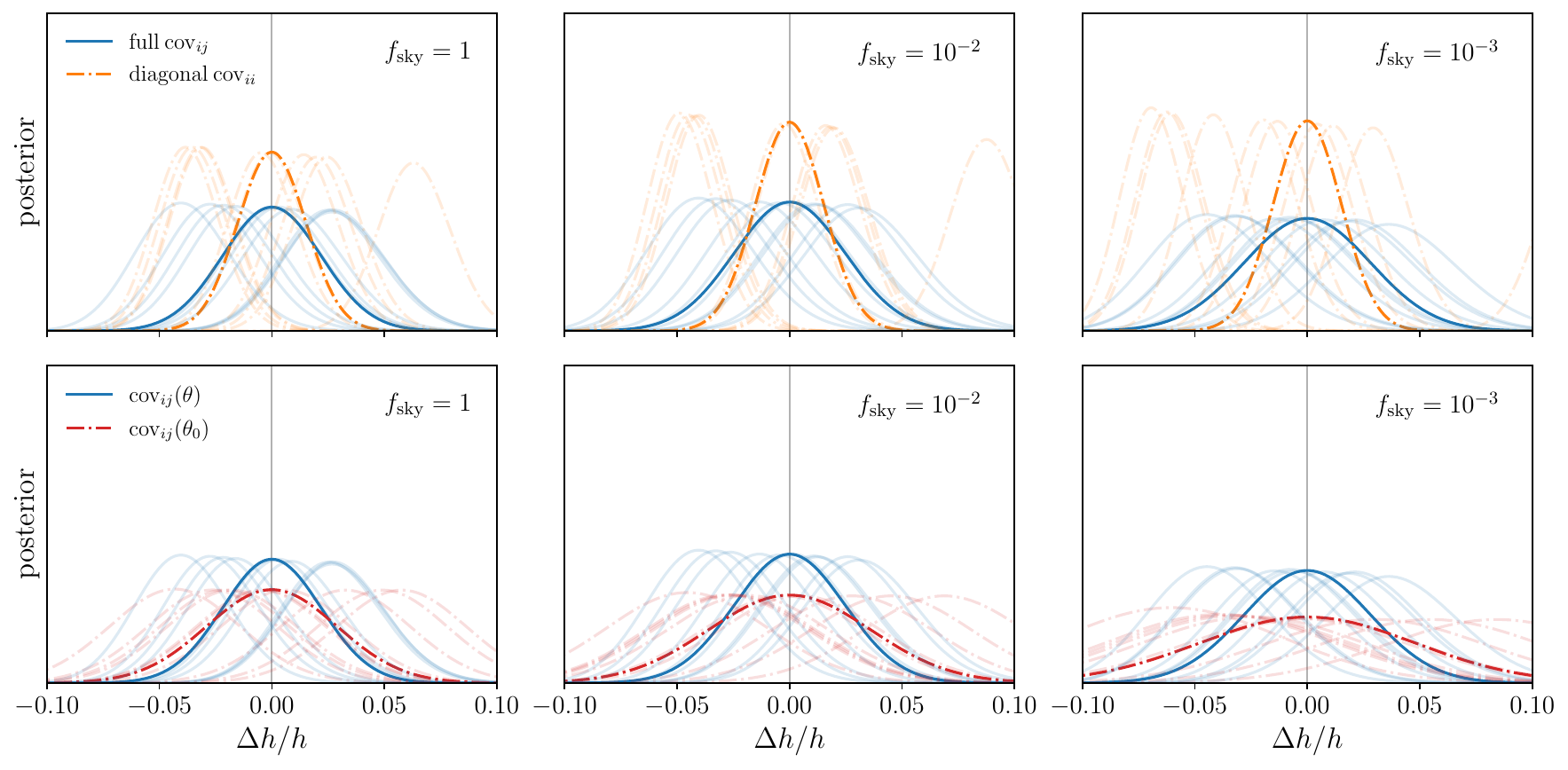}
    \caption{\textit{\textbf{Upper panels}}: Posterior distribution for various samples of 500 FRBs drawn with the respective covariance plotted in \Cref{fig:covariance_forecast}. Solid blue lines use the full covariance, while dashed orange lines treat the FRBs as independent and just use the diagonal of the covariance matrix and underestimate the true error severely by $40 \%$, $60 \%$ and $85 \%$ for the respective panels. The $x$-axis shows the relative deviation from the fiducial value (used to generate the synthetic data). Thick lines denote the average effect over many realisations, and shaded lines show different realisations of the noisy data. Single realisations analysed using diagonal covariance can lead to false parameter estimations. \textit{\textbf{Lower panels}}: Posterior distributions either using a parameter-dependent covariance (solid blue) or a fixed covariance (dashed red). \revision{Here the covariance is proportional to $h^2$ at leading order as can be seen from \cref{eq:DM_LSS_avg}. In other analysis, however, e.g. \citep{hagstotz_new_2022}, a prior is set on $\omega_\mathrm{b0} = \Omega_\mathrm{b0}h^2$, this would change the scaling of the covariance.} 
    The cosmological dependence of the covariance matrix contains additional information, shrinking the error bars by $30\%$, $45 \%$ and $70\%$ for the respective sky fractions compared to a covariance calculated at fixed parameters.}
    \label{fig:frb_forecast_posterior}
\end{figure}

In \Cref{fig:covariance_forecast} we show the correlation coefficient $r_{ij} = \mathrm{cov}_{ij}/(\mathrm{cov}_{ii}\mathrm{cov}_{jj})^{1/2}$ for 500 FRBs distributed over different parts of the sky. While the covariance for a few hundred events distributed over the full sphere is dominated by the diagonal elements, the same number of FRBs distributed on a small fraction of the sky leads to a tight correlation due to the small angular separation of events.

The full covariance modelling is crucial for parameter estimation from larger FRB catalogs. In \Cref{fig:frb_forecast_posterior} we show the posterior of $h$ from several synthetic catalogues of 500 events distributed over various fractions of the sky. The catalogue is always generated using the true covariance matrix, and analysed using either the full covariance (blue solid) or only the diagonal (assuming uncorrelated events, orange dashed). Thick lines are showing the average over many realisations, while single realisations of the data and the corresponding inference are shown with shaded lines. The assumption of uncorrelated DMs leads to a severe underestimation of the error by $40 \%$, $60 \%$ and up to $85 \%$ for events covering either the full sky, or $f_\mathrm{sky} = 10^{-2}$ and $f_\mathrm{sky} = 10^{-3}$ respectively. While a linear parameter cannot be biased on average, single realisations using the diagonal correlation matrix can easily show more than $3 \sigma$ deviation from the true value used to generate the samples.

In the lower panels of \Cref{fig:frb_forecast_posterior} we show the effect of the additional cosmological information contained in the covariance of the samples. We compare again inference using the full, parameter-dependent covariance matrix (solid blue) with the case of a fixed covariance matrix (dashed red). The width of the posterior shrinks by $30 \%$, $45 \%$ and up to $70 \%$ depending on the sky fraction.

\begin{figure}
    \centering
    \includegraphics[width = 1\textwidth]{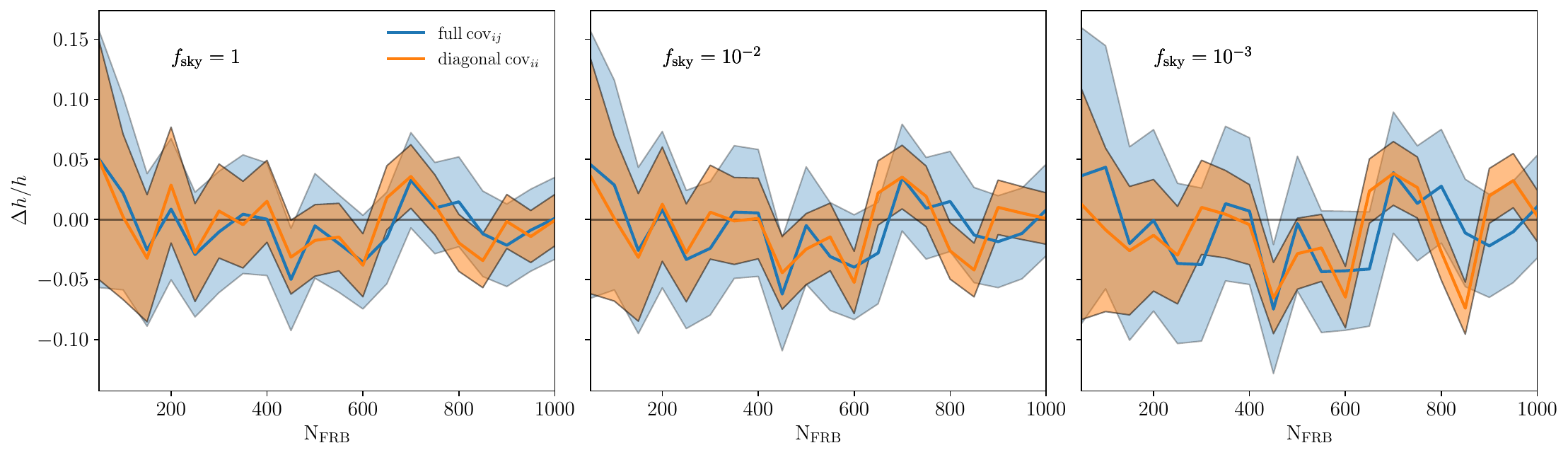}
    \caption{Best fit values and 95$\%$ confidence interval (shaded bands) against the number of FRBs with host identification generated with a known redshift distribution for a full sky sample, $f_\mathrm{sky}= 1$ and $f_\mathrm{sky}= 10^{-2}$, $f_\mathrm{sky}= 10^{-3}$ from left to right. Results from synthetic data analysed either using only the diagonal covariance (orange) or the full covariance including off-diagonal elements (blue). The diagonal covariance underestimates the true error, so the inferred value of $h$ is offset from the fiducial value.}
    \label{fig:bands}
\end{figure}

In \Cref{fig:bands} we demonstrate the influence of the covariance as a function of the number of observed FRBs, again for the same sky fractions. Note that the synthetic data used in \Cref{fig:frb_forecast_posterior} is not necessarily the same as in \Cref{fig:bands}, but both are compatible with the full covariance. The solid line shows the maximum posterior values while the shaded areas correspond to the 95\% confidence interval. From the plots it is noticeable that the uncertainty on $h$ is severely underestimated for $N_\mathrm{FRB} \geq 300$ even for a full sky sample when using a diagonal covariance. Although significant biases are unlikely to arise in this scenario, $3\sigma$ deviations from the true underlying value are possible if the covariance between events is neglected. For $f_\mathrm{sky} = 10^{-3}$ these effects are already present for smaller $N_\mathrm{FRB}$ and the error can be misestimated by up to 50 per-cent for $N_\mathrm{FRB}$ as low as 40. While the last case is mostly of academical nature, selecting subsets of close by FRBs which are close by and ignoring there covariance might be dangerous. 

We close this section with a short comparison with the approach used in e.g. \citet{macquart_census_2020},\citet{wu_8_2022} or \citet{james_measurement_2022}. These works use a likelihood derived from the one-point probability distribution function of $\mathrm{DM}_\mathrm{LSS}$ and take into account the full non-Gaussianity of the DM distribution since it is measured directly from numerical simulations. While this captures the high DM tail of the distribution, the final likelihood is still dominated by the variance rather than its skewness. On the other hand, it then is generally difficult to take the covariance between different FRBs into account since in principle an $N_\mathrm{FRB}$-point function is needed to obtain the accurate shape of the likelihood. Measuring all necessary moments from numerical simulations is inherently difficult due to the high noise in these estimates. Furthermore, it is challenging to include the parameter dependence in these approaches, since the numerical simulations are only evaluated at a single cosmology, although some of it has already been taken care of by looking at the relative DM, i.e. compared to the background cosmology. It would be interesting which effect is more important: the correlation or the high DM tail of the distribution. We refer this investigation to future work.

\section{Conclusions}
\label{sec:conclusion}
In this paper we have investigated the impact of the LSS induced correlation between FRBs with host identification. We have derived the covariance matrix in harmonic and real space for FRBs observed at redshift $z$ and $\hat{\boldsymbol{x}}_i$ position. This new covariance matrix was then used to reanalyse the FRBs from the FRB catalogue \citep{petroff_frbcat_2016} and to explore the influence on a single parameter, the Hubble constant $h$, measured from current and future samples. Our main findings can be summarised as follows:
\begin{enumerate}
    \item The number of current FRBs with host identification does not require the inclusion of the covariance between them as the statistical significance of the measurement is too low. Here we find similar results as \citet{hagstotz_new_2022}.
    
    \item For a full sky sample we find that the Hubble constant $h$ or any other linear model parameter picks up an underestimated error of roughly 50 per-cent for 500 FRBs in the best case. In the worst case there can be significant biases for any single realisation of the data. This situation becomes even more serious if the number of FRBs increases. 

    \item If the parameter dependence of the covariance is not accounted for, biases can arise already for smaller numbers of FRBs in the case of partial sky fraction. We generally advise to take the dependence on the model parameters of the covariance (diagonal or not) into account, as it contains complementary information to the background dispersion measure. 

    \item When small patches of the sky are observed ($f_\mathrm{sky} = 10^{-3}$ or smaller) the influence of the full covariance can be seen already for $N_\mathrm{FRB} = 40$, leading to underestimated errors. 
\end{enumerate}
We therefore conclude that the LSS covariance matrix of the DM of FRBs with host identification can become important 
in the future when more such FRBs ($\sim 10^2$) have been observed. Here we only investigated isotropically distributed FRB samples over sky patches of different sizes. In case of a more complex selection function the results found here might become more severe, but we leave this for future work. Another issue is the inclusion of the non-Gaussian structure of the likelihood which in principle is naturally included in approaches using a formula fitted to numerical simulations \citep{macquart_census_2020,wu_8_2022,james_measurement_2022}. These, however, lack the possibility to account for the correlations between the different FRBs. This approach is feasible at the moment, but will lead to errorneous conclusions in the future. \revision{Lastly, a lot of studies use multiple thousand FRBs to forecast the possible constraints \citep[see][for an extensive review on the possible science goals]{petroff_fast_2022}. Up to date no  forecast has accounted for correlations between different FRBs, hence overestimating the achievable precision of future experiments.
Furthermore,
there are studies investigating the possibility to constrain reionization with FRBs \citep{Heimersheim_what_2022}. These studies, due to their high redshift FRBs would be much stronger affected by the covariance matrix due to the longer integration path. Another field of application of the results presented here would be the afterglow of Gamma Ray Bursts (GRBs) which can also be used to trace the electron distribution in the Universe \citep{2003ApJ...598L..79I,2004MNRAS.348..999I}. Due to their large redshifts, one would expect the correlations of nearby pairs of GRBs to be substantial.}

\vspace{2mm}

\section*{Acknowledgments}
The authors would like to thank an anonymous referee for very valuable comments which helped to improve the manuscript and its implications.
RR is supported by the European Research Council (Grant No. 770935). SH was supported by
the Excellence Cluster ORIGINS which is funded by the Deutsche Forschungsgemeinschaft (DFG,
German Research Foundation) under Germany’s Excellence Strategy - EXC-2094 - 390783311. SH and RR acknowledge support by Institut Pascal at Université Paris-Saclay during the Paris-Saclay Astroparticle Symposium 2022, with the support of the P2IO Laboratory of Excellence (program “Investissements d’avenir” ANR-11-IDEX-0003-01 Paris-Saclay and ANR-10-LABX-0038), the P2I axis of the Graduate School of Physics of Université Paris-Saclay, as well as IJCLab, CEA, APPEC, IAS, OSUPS, and the IN2P3 master project UCMN.
\\ \\
{\bf Data Availability}: The data and code underlying this article will be shared on request to the corresponding author. 

\label{lastpage}
\bibliographystyle{mnras}
\bibliography{MyLibrary}

\begin{thebibliography}{}
\makeatletter
\relax
\def\mn@urlcharsother{\let\do\@makeother \do\$\do\&\do\#\do\^\do\_\do\%\do\~}
\def\mn@doi{\begingroup\mn@urlcharsother \@ifnextchar [ {\mn@doi@}
  {\mn@doi@[]}}
\def\mn@doi@[#1]#2{\def\@tempa{#1}\ifx\@tempa\@empty \href
  {http://dx.doi.org/#2} {doi:#2}\else \href {http://dx.doi.org/#2} {#1}\fi
  \endgroup}
\def\mn@eprint#1#2{\mn@eprint@#1:#2::\@nil}
\def\mn@eprint@arXiv#1{\href {http://arxiv.org/abs/#1} {{\tt arXiv:#1}}}
\def\mn@eprint@dblp#1{\href {http://dblp.uni-trier.de/rec/bibtex/#1.xml}
  {dblp:#1}}
\def\mn@eprint@#1:#2:#3:#4\@nil{\def\@tempa {#1}\def\@tempb {#2}\def\@tempc
  {#3}\ifx \@tempc \@empty \let \@tempc \@tempb \let \@tempb \@tempa \fi \ifx
  \@tempb \@empty \def\@tempb {arXiv}\fi \@ifundefined
  {mn@eprint@\@tempb}{\@tempb:\@tempc}{\expandafter \expandafter \csname
  mn@eprint@\@tempb\endcsname \expandafter{\@tempc}}}

\bibitem[\protect\citeauthoryear{Abbott et~al.}{Abbott
  et~al.}{2022}]{abbott_dark_2022}
Abbott T. M.~C.,  et~al., 2022, \mn@doi [Phys. Rev. D]
  {10.1103/PhysRevD.105.023520}, 105, 023520

\bibitem[\protect\citeauthoryear{Aghanim et~al.}{Aghanim
  et~al.}{2020a}]{Aghanim:2018eyx}
Aghanim N.,  et~al., 2020a, \mn@doi [Astron. Astrophys.]
  {10.1051/0004-6361/201833910}, 641, A6

\bibitem[\protect\citeauthoryear{Aghanim et~al.}{Aghanim
  et~al.}{2020b}]{aghanim_planck_2020}
Aghanim N.,  et~al., 2020b, \mn@doi [Astron. Astrophys.]
  {10.1051/0004-6361/201833910}, 641, A6

\bibitem[\protect\citeauthoryear{Andersen et~al.,}{Andersen
  et~al.}{2020}]{andersen_bright_2020}
Andersen B.,  et~al., 2020, \mn@doi [Nature] {10.1038/s41586-020-2863-y}, 587,
  54

\bibitem[\protect\citeauthoryear{Asgari et~al.}{Asgari
  et~al.}{2021}]{asgari_kids-1000_2021}
Asgari M.,  et~al., 2021, \mn@doi [Astron. Astrophys.]
  {10.1051/0004-6361/202039070}, 645, A104

\bibitem[\protect\citeauthoryear{Aver, Olive  \& Skillman}{Aver
  et~al.}{2015}]{Aver:2015iza}
Aver E.,  Olive K.~A.,   Skillman E.~D.,  2015, \mn@doi [JCAP]
  {10.1088/1475-7516/2015/07/011}, 07, 011

\bibitem[\protect\citeauthoryear{Bhattacharya, Kumar  \& Linder}{Bhattacharya
  et~al.}{2021}]{bhattacharya_fast_2020}
Bhattacharya M.,  Kumar P.,   Linder E.~V.,  2021, \mn@doi [Phys. Rev. D]
  {10.1103/PhysRevD.103.103526}, 103, 103526

\bibitem[\protect\citeauthoryear{Brout et~al.}{Brout
  et~al.}{2022}]{Brout:2022vxf}
Brout D.,  et~al., 2022, \mn@doi [Astrophys. J.] {10.3847/1538-4357/ac8e04},
  938, 110

\bibitem[\protect\citeauthoryear{Carron}{Carron}{2013}]{carron_assumption_2013}
Carron J.,  2013, \mn@doi [Astronomy \& Astrophysics]
  {10.1051/0004-6361/201220538}, 551, A88

\bibitem[\protect\citeauthoryear{Champion et~al.,}{Champion
  et~al.}{2016}]{champion_five_2016}
Champion D.~J.,  et~al., 2016, \mn@doi [Monthly Notices of the Royal
  Astronomical Society] {10.1093/mnrasl/slw069}, 460, L30

\bibitem[\protect\citeauthoryear{Chatterjee et~al.,}{Chatterjee
  et~al.}{2017}]{chatterjee_direct_2017}
Chatterjee S.,  et~al., 2017, \mn@doi [Nature] {10.1038/nature20797}, 541, 58

\bibitem[\protect\citeauthoryear{Connor, Sievers  \& Pen}{Connor
  et~al.}{2016}]{connor_non-cosmological_2016}
Connor L.,  Sievers J.,   Pen U.-L.,  2016, \mn@doi [Monthly Notices of the
  Royal Astronomical Society] {10.1093/mnrasl/slv124}, 458, L19

\bibitem[\protect\citeauthoryear{{Cooke}, {Pettini}  \& {Steidel}}{{Cooke}
  et~al.}{2018}]{cooke_2018_bbn}
{Cooke} R.~J.,  {Pettini} M.,   {Steidel} C.~C.,  2018, \mn@doi [\apj]
  {10.3847/1538-4357/aaab53}, \href
  {https://ui.adsabs.harvard.edu/abs/2018ApJ...855..102C} {855, 102}

\bibitem[\protect\citeauthoryear{{Deng} \& {Zhang}}{{Deng} \&
  {Zhang}}{2014}]{2014ApJ...783L..35D}
{Deng} W.,  {Zhang} B.,  2014, \mn@doi [\apjl] {10.1088/2041-8205/783/2/L35},
  \href {https://ui.adsabs.harvard.edu/abs/2014ApJ...783L..35D} {783, L35}

\bibitem[\protect\citeauthoryear{Fukugita \& Peebles}{Fukugita \&
  Peebles}{2004}]{Fukugita:2004ee}
Fukugita M.,  Peebles P. J.~E.,  2004, \mn@doi [Astrophys. J.]
  {10.1086/425155}, 616, 643

\bibitem[\protect\citeauthoryear{Hagstotz, Reischke  \& Lilow}{Hagstotz
  et~al.}{2022}]{hagstotz_new_2022}
Hagstotz S.,  Reischke R.,   Lilow R.,  2022, \mn@doi [Mon. Not. Roy. Astron.
  Soc.] {10.1093/mnras/stac077}, 511, 662

\bibitem[\protect\citeauthoryear{Heimersheim, Sartorio, Fialkov  \&
  Lorimer}{Heimersheim et~al.}{2022}]{Heimersheim_what_2022}
Heimersheim S.,  Sartorio N.~S.,  Fialkov A.,   Lorimer D.~R.,  2022, \mn@doi
  [The Astrophysical Journal] {10.3847/1538-4357/ac70c9}, 933, 57

\bibitem[\protect\citeauthoryear{{Inoue}}{{Inoue}}{2004}]{2004MNRAS.348..999I}
{Inoue} S.,  2004, \mn@doi [\mnras] {10.1111/j.1365-2966.2004.07359.x}, \href
  {https://ui.adsabs.harvard.edu/abs/2004MNRAS.348..999I} {348, 999}

\bibitem[\protect\citeauthoryear{{Ioka}}{{Ioka}}{2003}]{2003ApJ...598L..79I}
{Ioka} K.,  2003, \mn@doi [\apjl] {10.1086/380598}, \href
  {https://ui.adsabs.harvard.edu/abs/2003ApJ...598L..79I} {598, L79}

\bibitem[\protect\citeauthoryear{James et~al.,}{James
  et~al.}{2022}]{james_measurement_2022}
James C.~W.,  et~al., 2022, \mn@doi [Monthly Notices of the Royal Astronomical
  Society] {10.1093/mnras/stac2524}, 516, 4862

\bibitem[\protect\citeauthoryear{Kitching \& Heavens}{Kitching \&
  Heavens}{2017}]{Kitching:2016xcl}
Kitching T.~D.,  Heavens A.~F.,  2017, \mn@doi [Phys. Rev. D]
  {10.1103/PhysRevD.95.063522}, 95, 063522

\bibitem[\protect\citeauthoryear{Kodwani, Alonso  \& Ferreira}{Kodwani
  et~al.}{2019}]{kodwani_effect_2019}
Kodwani D.,  Alonso D.,   Ferreira P.,  2019, \mn@doi [The Open Journal of
  Astrophysics] {10.21105/astro.1811.11584}, 2, 10.21105/astro.1811.11584

\bibitem[\protect\citeauthoryear{Limber}{Limber}{1953}]{limber_analysis_1953}
Limber D.~N.,  1953, \mn@doi [\apj] {10.1086/145672}, 117, 134

\bibitem[\protect\citeauthoryear{Loverde \& Afshordi}{Loverde \&
  Afshordi}{2008}]{loverde_extended_2008}
Loverde M.,  Afshordi N.,  2008, \mn@doi [\prd] {10.1103/PhysRevD.78.123506},
  78, 123506

\bibitem[\protect\citeauthoryear{Macquart et~al.,}{Macquart
  et~al.}{2020}]{macquart_census_2020}
Macquart J.-P.,  et~al., 2020, \mn@doi [Nature] {10.1038/s41586-020-2300-2},
  581, 391

\bibitem[\protect\citeauthoryear{Madau \& Dickinson}{Madau \&
  Dickinson}{2014}]{Madau:2014bja}
Madau P.,  Dickinson M.,  2014, \mn@doi [Ann. Rev. Astron. Astrophys.]
  {10.1146/annurev-astro-081811-125615}, 52, 415

\bibitem[\protect\citeauthoryear{Masui \& Sigurdson}{Masui \&
  Sigurdson}{2015}]{masui_dispersion_2015}
Masui K.~W.,  Sigurdson K.,  2015, \mn@doi [Physical Review Letters]
  {10.1103/PhysRevLett.115.121301}, 115, 121301

\bibitem[\protect\citeauthoryear{McCarthy, Bird, Schaye, Harnois-Deraps, Font
  \& van Waerbeke}{McCarthy et~al.}{2018}]{mccarthy_bahamas_2018}
McCarthy I.~G.,  Bird S.,  Schaye J.,  Harnois-Deraps J.,  Font A.~S.,   van
  Waerbeke L.,  2018, \mn@doi [Monthly Notices of the Royal Astronomical
  Society] {10.1093/mnras/sty377}, 476, 2999

\bibitem[\protect\citeauthoryear{McQuinn}{McQuinn}{2014}]{mcquinn_locating_2014}
McQuinn M.,  2014, \mn@doi [The Astrophysical Journal Letters]
  {10.1088/2041-8205/780/2/L33}, 780, L33

\bibitem[\protect\citeauthoryear{Mead, Peacock, Heymans, Joudaki  \&
  Heavens}{Mead et~al.}{2015}]{mead_accurate_2015}
Mead A.~J.,  Peacock J.~A.,  Heymans C.,  Joudaki S.,   Heavens A.~F.,  2015,
  \mn@doi [\mnras] {10.1093/mnras/stv2036}, 454, 1958

\bibitem[\protect\citeauthoryear{Mead, Tr\"oster, Heymans, Van~Waerbeke  \&
  McCarthy}{Mead et~al.}{2020}]{mead_hydrodynamical_2020}
Mead A.~J.,  Tr\"oster T.,  Heymans C.,  Van~Waerbeke L.,   McCarthy I.~G.,
  2020, \mn@doi [Astron. Astrophys.] {10.1051/0004-6361/202038308}, 641, A130

\bibitem[\protect\citeauthoryear{Petroff et~al.,}{Petroff
  et~al.}{2015}]{petroff_real-time_2015}
Petroff E.,  et~al., 2015, \mn@doi [MNRAS] {10.1093/mnras/stu2419}, 447, 246

\bibitem[\protect\citeauthoryear{Petroff et~al.,}{Petroff
  et~al.}{2016}]{petroff_frbcat_2016}
Petroff E.,  et~al., 2016, \mn@doi [Publications of the Astronomical Society of
  Australia] {10.1017/pasa.2016.35}, 33, e045

\bibitem[\protect\citeauthoryear{Petroff, Hessels  \& Lorimer}{Petroff
  et~al.}{2022}]{petroff_fast_2022}
Petroff E.,  Hessels J. W.~T.,   Lorimer D.~R.,  2022, \mn@doi [Astron.
  Astrophys. Rev.] {10.1007/s00159-022-00139-w}, 30, 2

\bibitem[\protect\citeauthoryear{Rafiei-Ravandi, Smith  \&
  Masui}{Rafiei-Ravandi et~al.}{2020}]{rafiei-ravandi_characterizing_2020}
Rafiei-Ravandi M.,  Smith K.~M.,   Masui K.~W.,  2020, \mn@doi [Phys. Rev. D]
  {10.1103/PhysRevD.102.023528}, 102, 023528

\bibitem[\protect\citeauthoryear{Rafiei-Ravandi et~al.}{Rafiei-Ravandi
  et~al.}{2021}]{rafiei-ravandi_chimefrb_2021}
Rafiei-Ravandi M.,  et~al., 2021, \mn@doi [Astrophys. J.]
  {10.3847/1538-4357/ac1dab}, 922, 42

\bibitem[\protect\citeauthoryear{Reischke, Kiessling  \& Schäfer}{Reischke
  et~al.}{2017}]{reischke_variations_2017}
Reischke R.,  Kiessling A.,   Schäfer B.~M.,  2017, \mn@doi [\mnras]
  {10.1093/mnras/stw2976}, 465, 4016

\bibitem[\protect\citeauthoryear{Reischke, Hagstotz  \& Lilow}{Reischke
  et~al.}{2021}]{reischke_probing_2021}
Reischke R.,  Hagstotz S.,   Lilow R.,  2021, \mn@doi [Phys. Rev. D]
  {10.1103/PhysRevD.103.023517}, 103, 023517

\bibitem[\protect\citeauthoryear{Reischke, Hagstotz  \& Lilow}{Reischke
  et~al.}{2022}]{reischke_consistent_2022}
Reischke R.,  Hagstotz S.,   Lilow R.,  2022, Monthly Notices of the Royal
  Astronomical Society, 512, 285

\bibitem[\protect\citeauthoryear{Riess et~al.}{Riess
  et~al.}{2022}]{riess_comprehensive_2022}
Riess A.~G.,  et~al., 2022, \mn@doi [Astrophys. J. Lett.]
  {10.3847/2041-8213/ac5c5b}, 934, L7

\bibitem[\protect\citeauthoryear{Shirasaki, Kashiyama  \& Yoshida}{Shirasaki
  et~al.}{2017}]{shirasaki_large-scale_2017}
Shirasaki M.,  Kashiyama K.,   Yoshida N.,  2017, \mn@doi [Physical Review D]
  {10.1103/PhysRevD.95.083012}, 95, 083012

\bibitem[\protect\citeauthoryear{Takahashi, Ioka, Mori  \& Funahashi}{Takahashi
  et~al.}{2021}]{takahashi_statistical_2021}
Takahashi R.,  Ioka K.,  Mori A.,   Funahashi K.,  2021, \mn@doi [Mon. Not.
  Roy. Astron. Soc.] {10.1093/mnras/stab170}, 502, 2615

\bibitem[\protect\citeauthoryear{Thornton et~al.,}{Thornton
  et~al.}{2013}]{thornton_population_2013}
Thornton D.,  et~al., 2013, \mn@doi [Science] {10.1126/science.1236789}, 341,
  53

\bibitem[\protect\citeauthoryear{Tröster et~al.}{Tröster
  et~al.}{2022}]{troster_joint_2022}
Tröster T.,  et~al., 2022, \mn@doi [Astron. Astrophys.]
  {10.1051/0004-6361/202142197}, 660, A27

\bibitem[\protect\citeauthoryear{Walters, Weltman, Gaensler, Ma  \&
  Witzemann}{Walters et~al.}{2018}]{walters_future_2018}
Walters A.,  Weltman A.,  Gaensler B.~M.,  Ma Y.-Z.,   Witzemann A.,  2018,
  \mn@doi [The Astrophysical Journal] {10.3847/1538-4357/aaaf6b}, 856, 65

\bibitem[\protect\citeauthoryear{Wu, Zhang  \& Wang}{Wu
  et~al.}{2022}]{wu_8_2022}
Wu Q.,  Zhang G.-Q.,   Wang F.-Y.,  2022, \mn@doi [Monthly Notices of the Royal
  Astronomical Society] {10.1093/mnrasl/slac022}, 515, L1

\bibitem[\protect\citeauthoryear{Zhang, Yan, Li, Zhang  \& Wang}{Zhang
  et~al.}{2021}]{zhang_intergalactic_2020}
Zhang Z.~J.,  Yan K.,  Li C.~M.,  Zhang G.~Q.,   Wang F.~Y.,  2021, \mn@doi
  [Astrophys. J.] {10.3847/1538-4357/abceb9}, 906, 49

\bibitem[\protect\citeauthoryear{Zhou, Li, Wang, Fan  \& Wei}{Zhou
  et~al.}{2014}]{zhou_fast_2014}
Zhou B.,  Li X.,  Wang T.,  Fan Y.-Z.,   Wei D.-M.,  2014, \mn@doi [Physical
  Review D] {10.1103/PhysRevD.89.107303}, 89, 107303

\bibitem[\protect\citeauthoryear{de~la Bella, Tessore  \& Bridle}{de~la Bella
  et~al.}{2020}]{de_la_bella_unequal-time_2020}
de~la Bella L.~F.,  Tessore N.,   Bridle S.,  2020, arXiv:2011.06185 [astro-ph]

\makeatother
\end{thebibliography}
\end{document}